\documentclass[twocolumn]{aastex62}

\usepackage{amsmath}
\usepackage{xcolor}
\usepackage[multiple]{footmisc}

\newcommand{\real}{{\text{Re}}}
\newcommand{\imag}{{\text{Im}}}
\newcommand{\Tr}{{\text{Tr}}}
\newcommand{\pNL}{p_{\text{NL}}}
\newcommand{\piL}{\pi_{\text{L}}}
\newcommand{\piNL}{\pi_{\text{NL}}}
\newcommand{\E}{{\text{E}}}
\newcommand{\Var}{{\text{Var}}}
\newcommand{\Cov}{{\text{Cov}}}

\newcommand{\mL}{{\mathcal{L}}}
\newcommand{\mLeff}{{\mathcal{L}_{\text{eff}}}}
\newcommand{\mM}{{\mathcal{M}}}
\newcommand{\bmM}{{\boldsymbol{\mathcal{M}}}}
\newcommand{\mN}{{\mathcal{N}}}
\newcommand{\mCN}{{\mathcal{CN}}}
\newcommand{\ba}{{\boldsymbol{a}}}
\newcommand{\bA}{{\boldsymbol{A}}}
\newcommand{\bb}{{\boldsymbol{b}}}
\newcommand{\bB}{{\boldsymbol{B}}}
\newcommand{\bF}{{\boldsymbol{F}}}
\newcommand{\bj}{{\boldsymbol{j}}}
\newcommand{\bs}{{\boldsymbol{s}}}
\newcommand{\bS}{{\boldsymbol{S}}}
\newcommand{\bI}{{\boldsymbol{I}}}
\newcommand{\boldv}{{\boldsymbol{v}}}

\newcommand{\bZ}{{\boldsymbol{Z}}}

\newcommand{\bLambda}{{\boldsymbol{\Lambda}}}
\newcommand{\btheta}{{\boldsymbol{\theta}}}
\newcommand{\bzeta}{{\boldsymbol{\zeta}}}
\newcommand{\brho}{{\boldsymbol{\rho}}}
\newcommand{\bepsilon}{{\boldsymbol{\varepsilon}}}
\newcommand{\bthetaL}{{\boldsymbol{\theta}_{\text{L}}}}
\newcommand{\bthetaNL}{{\boldsymbol{\theta}_{\text{NL}}}}
\newcommand{\bC}{{\boldsymbol{C}}}
\newcommand{\boldm}{{\boldsymbol{m}}}
\newcommand{\by}{{\boldsymbol{y}}}
\newcommand{\balpha}{{\boldsymbol{\alpha}}}
\newcommand{\bPhi}{{\boldsymbol{\Phi}}}
\newcommand{\bXi}{{\boldsymbol{\Xi}}}

\newcommand{\bmuL}{{\boldsymbol{\mu}_{\text{L}}}}
\newcommand{\bSigmaL}{{\boldsymbol{\Sigma}_{\text{L}}}}

\shorttitle{Global 21-cm signal extraction IV: including receiver uncertainties}
\shortauthors{Tauscher et al.}

\begin{document}

\title{Global 21-cm signal extraction from foreground and instrumental effects IV:

Accounting for realistic instrument uncertainties and their overlap with foreground and signal models}

\correspondingauthor{Keith Tauscher}

\author{Keith~Tauscher}
\affiliation{Center for Astrophysics and Space Astronomy, Department of Astrophysical and Planetary Science, University of Colorado, Boulder, CO 80309, USA}

\author{David~Rapetti}
\affiliation{NASA Ames Research Center, Moffett Field, CA 94035, USA}
\affiliation{Research Institute for Advanced Computer Science, Universities Space Research Association, Mountain View, CA 94043, USA}
\affiliation{Center for Astrophysics and Space Astronomy, Department of Astrophysical and Planetary Science, University of Colorado, Boulder, CO 80309, USA}

\author{Bang~D.~Nhan}
\affiliation{Central Development Laboratory, National Radio Astronomy Observatory, Charlottesville, VA 22903, USA}
\affiliation{Department of Astronomy, University of Virginia, Charlottesville, VA 22903, USA}

\author{Alec~Handy}
\affiliation{Central Development Laboratory, National Radio Astronomy Observatory, Charlottesville, VA 22903, USA}

\author{Neil~Bassett}
\affiliation{Center for Astrophysics and Space Astronomy, Department of Astrophysical and Planetary Science, University of Colorado, Boulder, CO 80309, USA}

\author{Joshua~Hibbard}
\affiliation{Center for Astrophysics and Space Astronomy, Department of Astrophysical and Planetary Science, University of Colorado, Boulder, CO 80309, USA}

\author{David~Bordenave}
\affiliation{Department of Astronomy, University of Virginia, Charlottesville, VA 22903, USA}

\author{Richard~F.~Bradley}
\affiliation{Central Development Laboratory, National Radio Astronomy Observatory, Charlottesville, VA 22903, USA}

\author{Jack~O.~Burns}
\affiliation{Center for Astrophysics and Space Astronomy, Department of Astrophysical and Planetary Science, University of Colorado, Boulder, CO 80309, USA}

\email{Keith.Tauscher@colorado.edu}

\begin{abstract}
  All 21-cm signal experiments rely on electronic receivers that affect the data via both multiplicative and additive biases through the receiver's gain and noise temperature.~While experiments attempt to remove these biases, the residuals of their imperfect calibration techniques can still confuse signal extraction algorithms. In this paper, the fourth and final installment of our pipeline series, we present a technique for fitting out receiver effects as efficiently as possible. The fact that the gain and global signal, which are multiplied in the observation equation, must both be modeled implies that the model of the data is nonlinear in its parameters, making numerical sampling the only way to explore the parameter distribution rigorously.~However, multi-spectra fits, which are necessary to extract the signal confidently as demonstrated in the third paper of the series, often require large numbers of foreground parameters, increasing the dimension of the posterior distribution that must be explored and therefore causing numerical sampling inefficiencies. Building upon techniques in the second paper of the series, we outline a method to explore the full parameter distribution by numerically sampling a small subset of the parameters and analytically marginalizing over the others. We test this method in simulation using a type-I Chebyshev band-pass filter gain model and a fast signal model based on a spline between local extrema. The method works efficiently, converging quickly to the posterior signal parameter distribution. The final signal uncertainties are of the same order as the noise in the data.
\end{abstract}


\section{Introduction}
\label{sec:introduction}

The 21-cm line of neutral hydrogen has been used to see clouds of gas in the local universe for decades \citep{Ewen:51}, but in recent years, it has been theorized that highly redshifted emission and absorption from this line could track the history of the hydrogen gas in the early universe \citep[see, e.g.][]{Furlanetto:06,Morales:10,Pritchard:12,Loeb:13}. The postulated observable that allows this is known as the 21-cm signal and is essentially a perturbation on top of the cosmic microwave background (CMB), which is positive (negative) when the line is in emission (absorption) with respect to the CMB.

The 21-cm signal can be observed in two different ways, an angular power spectrum that characterizes spatial variations as a function of redshift \citep[as measured by, e.g.~HERA;][]{Deboer:17} and a sky-averaged spectrum known as the global signal that statistically describes the spatial mean behavior of the universe over time, which is the subject of this paper.~Multiple single-antenna experiments that are currently observing or are under active development are attempting to measure the global signal, such as the Large-aperture Experiment to detect the Dark Age \citep[LEDA;][]{Price:18}, Experiment to Detect the Global Epoch of Reionization (EoR) Signature \citep[EDGES;][]{Bowman:18,Monsalve:19,Mahesh:21}, Shaped Antenna measurement of the background RAdio Spectrum \citep[SARAS;][]{Singh:17}, Radio Experiment for the Analysis of Cosmic Hydrogen \citep[REACH;][]{Eloy:19,Anstey:20,Shen:21}, the Probing Radio Intensity at high-$z$ from Marion \citep[PRIzM;][]{Philip:19}, the Cosmic Twilight Polarimeter \citep[CTP;][]{Nhan:17,Nhan:19}, and the space-based mission concept Dark Ages Polarimeter PathfindER  \citep[DAPPER;][]{Burns:17,Burns:21a,Burns:21b}.

As has been pointed out by many, the main problem that must be solved for the global 21-cm signal, which is on the order of a few hundred millikelvin, to be confidently quantified is its extraction from the large foreground emission, which is on the order of a few thousand kelvin \citep{Liu:13,Switzer:14,Vedantham:14,Anstey:20}.\footnote{For a more comprehensive review of techniques proposed to solve the foreground problem in the global signal context, see \cite{Tauscher:18}.}~This paper is the fourth and final installment in a series laying out a pipeline designed to solve this problem. In \cite{Tauscher:18} (hereafter referred to as Paper I), we put forth a method of using training sets of foreground and signal spectra and singular value decomposition to form models of the components instead of assuming analytical, a priori models like polynomials, which are unlikely to fit the true foreground in the presence of a typically chromatic beam used for global 21-cm experiments, as shown in \cite{Hibbard:20} and \cite{Tauscher:20b}. The methods of Paper I result in constraints on the signal in frequency-temperature space. In \cite{Rapetti:20} (hereafter referred to as Paper II), we extended the method to an efficient exploration of the full posterior parameter distribution using any chosen nonlinear signal model, resulting in parameter constraints and covariances. Paper II also presents an analytical marginalization technique that greatly reduces the number of parameters that must be explored in order to rigorously sample the posterior signal parameter distribution. In \cite{Tauscher:20a} (hereafter referred to as Paper III), we showed that, in order to achieve uncertainties allowing for confident detection of the global 21-cm signal, one must perform a fit that correlates and models many spectra simultaneously. This stands in contrast with the analysis methods of many current experiments, including EDGES \citep{Bowman:18}, which only fit individual time-averaged spectra.

In this paper, we will combine the methods of papers I-III as summarized above into a complete form that includes receiver uncertainties, which have been neglected until now.~We will lay out a pipeline that first obtains signal estimates in frequency-temperature space using a modeling technique similar to the one in Paper I and then uses those estimates to follow up with a fit using any chosen nonlinear signal model that employs an analytical marginalization technique very similar to the one described in Paper II. Throughout this paper, we will be fitting ten concatenated spectra that are simulated at different local sidereal times (LST) because, as found in Paper III, it is necessary to achieve uncertainties below the K level.

In Section~\ref{sec:AMLP}, we describe the general technique we use to explore posterior distributions with large numbers of parameters as efficiently as possible, which we term analytical marginalization of linear parameters (AMLP).
In Section~\ref{sec:fitting-strategy}, we lay out how we use the technique described in Section~\ref{sec:AMLP} to fit 21-cm global signal data. In Section~\ref{sec:models}, we specify the models used for the beam-weighted foreground, global 21-cm signal, and receiver gain and noise temperature. In Section~\ref{sec:results}, we show the results of our method in simulation. In Section~\ref{sec:discussion}, we discuss caveats and simplifications used in this proof-of-concept work and how we plan to handle them when fitting observed data. Finally, we conclude in Section~\ref{sec:conclusions}.

\section{Analytical marginalization of linear parameters (AMLP)} \label{sec:AMLP}

In this section, we describe our method of exploring certain classes of distributions with large numbers of parameters as efficiently as possible.~In particular, AMLP allows any distribution with one or more Gaussian conditional distributions to be explored much more efficiently.

\subsection{Form of joint posterior}

In the following, we assume a Gaussian likelihood function,
\begin{subequations}
\begin{align}
  \mL(\btheta) &= \Pr[\by|\btheta], \\
  &\propto \exp\left\{-\frac{1}{2}\left[\by-\boldm(\btheta)\right]^T\bC^{-1}\left[\by-\boldm(\btheta)\right]\right\}, \label{eq:gaussian-likelihood}
\end{align}
\end{subequations}
where $\by$ is the data being fit, $\boldm(\btheta)$ is the model of the data evaluated at parameters $\btheta$, and $\bC$ is the noise covariance of the data. Using Bayes' theorem we can find the posterior distribution on the parameters,
\begin{subequations}
\begin{align}
  p(\btheta) &= \Pr[\btheta|\by] \\
  &= \pi(\btheta)\ \mL(\btheta),
  \label{eq:full-posterior}
\end{align}
\end{subequations}
where $\pi(\btheta)$ is the prior distribution that quantifies our a priori knowledge of the parameters.

\subsection{Conditions of AMLP} \label{sec:conditions-of-AMLP}

AMLP relies on splitting $\btheta$ into two groups, $\bthetaNL$ (nonlinear parameters) and $\bthetaL$ (linear parameters),\footnote{The use of NL short for nonlinear and L short for linear will be consistent throughout. In the following, we will sometimes write $\btheta$ dependence as dependence on $\bthetaL$ and $\bthetaNL$, e.g. $\mL(\btheta)=\mL(\bthetaL,\bthetaNL)$.} such that the conditional distribution of $\bthetaL$ at constant $\bthetaNL$, $\Pr[\bthetaL|\by,\bthetaNL]$, is Gaussian. With a Gaussian likelihood as given in Equation~\ref{eq:gaussian-likelihood}, the following is a common set of sufficient conditions for this to be true:
\begin{enumerate}
  \item The model conditioned on $\bthetaNL$ is linear in $\bthetaL$, i.e. $\frac{\partial \boldm}{\partial\bthetaL}=\bA(\bthetaNL)$ where the gradient matrix $\bA$ is independent of $\bthetaL$ but in general depends on $\bthetaNL$.
  \item The priors on the two sets of parameters are independent, i.e. $\pi(\btheta)=\piL(\bthetaL)\ \piNL(\bthetaNL)$.
  \item $\piL(\bthetaL)$ is Gaussian.\footnote{The distribution $\piL(\bthetaL)$ could also be an improper uniform prior, i.e. $\piL(\bthetaL)=1$, which would imply that any value of $\bthetaL$ is equally likely.}
\end{enumerate}

\subsection{Efficient calculation of marginal posterior}

Under the conditions described in Section~\ref{sec:conditions-of-AMLP}, it is possible to characterize the distribution only through numerical sampling of $\bthetaNL$ instead of $\btheta$ as a whole. This can greatly reduce the number of dimensions being explored by the sampling algorithm, which often drastically reduces the time necessary to effectively converge to the posterior distribution. To do this, we must find the posterior marginal distribution of $\bthetaNL$, which is given by
\begin{subequations}
\begin{align}
  \pNL(\bthetaNL) &= \int p(\bthetaNL,\bthetaL)\  d\bthetaL, \\
  &= \piNL(\bthetaNL)\ \mLeff(\bthetaNL). \label{eq:marginal-posterior-intermediate}
\end{align}
\end{subequations}
Here, we have implicitly defined the effective likelihood function $\mLeff(\bthetaNL)$ as
\begin{equation}
  \mLeff(\bthetaNL) = \int \piL(\bthetaL)\ \mL(\bthetaNL,\bthetaL)\ d\bthetaL.
\end{equation}
This integral can be computed directly by noting that, under the conditions mentioned above, the integrand is proportional to a multivariate Gaussian distribution\footnote{In particular, the integrand is proportional to the conditional distribution $\Pr[\bthetaL|\by,\bthetaNL]$, which should be read as the probability density of the linear parameters, $\bthetaL$, conditioned on the data, $\by$, and the nonlinear parameters, $\bthetaNL$.} in $\bthetaL$ with mean $\bmuL(\bthetaNL)=\E[\bthetaL|\bthetaNL]$ and covariance $\bSigmaL(\bthetaNL)=\Cov[\bthetaL|\bthetaNL]$. The conditional mean $\bmuL(\bthetaNL)$ and covariance $\bSigmaL(\bthetaNL)$ can be computed quickly because the model is linear when conditioned on $\bthetaNL$. We find
\begin{multline}
  \mLeff(\bthetaNL) = \left\Vert 2\pi \bSigmaL(\bthetaNL)\right\Vert^{1/2}\\ \times \piL\big(\bmuL(\bthetaNL)\big)\ \mL\big(\bthetaNL,\bmuL(\bthetaNL)\big),
\end{multline}
where $\left\Vert\cdot\right\Vert$ indicates the determinant. Plugging this into Equation~\ref{eq:marginal-posterior-intermediate}, we can write
\begin{multline}
  \pNL(\bthetaNL) = \left\Vert 2\pi\bSigmaL(\bthetaNL)\right\Vert^{1/2}\ \piNL(\bthetaNL)\\\times \piL\big(\bmuL(\bthetaNL)\big)\ \mL\big(\bthetaNL,\bmuL(\bthetaNL)\big). \label{eq:marginal-posterior}
\end{multline}

\subsection{Recreating sample from joint posterior}

The distribution $\pNL$ from Equation~\ref{eq:marginal-posterior} can be sampled numerically with a method such as Markov Chain Monte Carlo (MCMC) or nested sampling to yield a sequence of values $\{\btheta_{\text{NL}}^{(1)},\btheta_{\text{NL}}^{(2)},\ldots,\btheta_{\text{NL}}^{(N)}\}$.\footnote{For this work, we use the \texttt{emcee} code described in \cite{DFM:13}, although we also introduced a custom Metropolis Hastings MCMC sampler in \cite{Rapetti:20} that can be found in the \texttt{pylinex} code \citep{pylinex:21}.} From this sequence, we can create a sample of $\bthetaL$ by sampling the Gaussian conditional distributions. For each integer $k$ satisfying $1\le k\le N$, we sample\footnote{In this paper, $\mN(\boldv,\bZ)$ and $\mCN(\boldv,\bZ)$ denote normal and complex normal distributions with mean $\boldv$ and covariance $\bZ$.}
\begin{equation}
  \btheta_{\text{L}}^{(k)}\sim \mN\bigg(\bmuL\big(\btheta_{\text{NL}}^{(k)}\big),\bSigmaL\big(\btheta_{\text{NL}}^{(k)}\big)\bigg). \label{eq:sampling-conditional-distribution}
\end{equation}
Then, if we define $\btheta^{(k)}$ as $\begin{bmatrix} \btheta_{\text{NL}}^{(k)} \\ \btheta_{\text{L}}^{(k)} \end{bmatrix}$, $\left\{ \btheta^{(1)}, \btheta^{(2)}, \ldots, \btheta^{(N)} \right\}$ is a sample from $p(\btheta)$.\footnote{Note that if we only want a sample of $\bthetaL$, then instead of sampling the distribution in Equation~\ref{eq:sampling-conditional-distribution} only once per $k$-value, we can sample it $M$ times, leading to a total sample of $N\times M$ realizations of $\bthetaL$.}

\subsection{Choosing how to split parameters}

One remaining question about the AMLP technique is how to choose which parameters to include in $\bthetaNL$ and which to include in $\bthetaL$. For example, if the data vector consists of the product of two components, $A$ and $B$, which each have linear models with parameter vectors $\bepsilon$ and $\brho$, respectively, then there are multiple ways of applying AMLP. For instance, $\begin{bmatrix} \bthetaNL \\ \bthetaL \end{bmatrix}$ can be set to either $\begin{bmatrix} \bepsilon \\ \brho \end{bmatrix}$ or $\begin{bmatrix} \brho \\ \bepsilon \end{bmatrix}$. To decide which of these choices is the best, we examine the dimensions of $\bepsilon$ and $\brho$. The longer of the two vectors should be marginalized over (i.e. set to $\bthetaL$), whereas the other should be explored numerically (i.e. set to $\bthetaNL$).

\subsection{Implementation in the \texttt{pylinex} code} \label{sec:pylinex-implementation}

Through the \texttt{MultiConditionalFitModel} class in the \texttt{pylinex} code \citep{pylinex:21}, we can perform AMLP with any model that can be written using a combination of products and sums of constituent submodels (such as the one described in Section~\ref{sec:full-model} and shown in Figure~\ref{fig:specific-model-tree}) and any split of parameters into $\bthetaNL$ and $\bthetaL$ that satisfies the condition that the model at constant $\bthetaNL$ is linear in $\bthetaL$.

\section{Fitting data with foreground, signal, and receiver} \label{sec:fitting-strategy}

In this section, we lay out the general framework we use to apply AMLP to 21-cm observations with receiver effects. Here, it is important to note that when we refer to linear and nonlinear, we are referring to parameters and not the receiver, which is assumed to be operated in a linear regime.

\subsection{Modeling observations} \label{sec:full-model}

\begin{figure}[t!]
  \centering
  \includegraphics[width=0.45\textwidth]{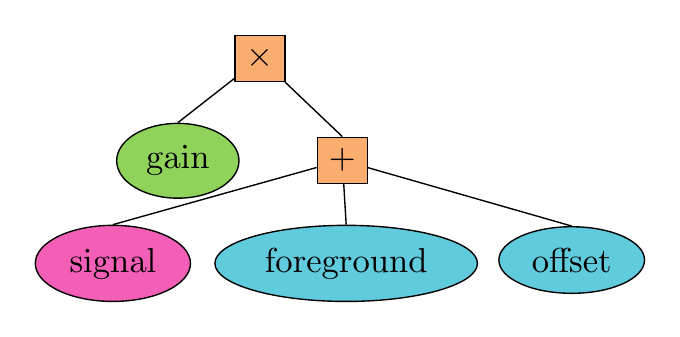}
  \caption{Visualization of the model tree corresponding to Equation~\ref{eq:block-model}. The ellipses (nodes with no children) represent the four components of the model, while the orange squares (branching nodes) represent their connections through either addition or multiplication. Blue ellipses indicate components (foreground and receiver offset) whose parameters will be marginalized at each step of an MCMC, i.e.~models whose parameters will be part of $\bthetaL$ as described in Section~\ref{sec:AMLP}. The green ellipse indicates that the gain parameters are included in $\bthetaNL$ and are explored via MCMC. The ellipse in magenta indicates a component (signal) whose parameters are marginalized over if they are linear, but explored via MCMC if they are nonlinear.} \label{fig:specific-model-tree}
\end{figure}

The full observation equation of the temperature data $T_{\nu,t}$ as a function of frequency $\nu$ and time $t$ is given by
\begin{equation}
  T_{\nu,t} = G_\nu\times(f_{\nu,t}+s_\nu+o_\nu) + n_{\nu,t},  \label{eq:observation-equation}
\end{equation}
where $G_\nu$ is the receiver's multiplicative bias (gain),\footnote{$G_\nu$ is the power gain, which is equal to $|g_\nu|^2$, where $g_\nu$ is the complex voltage gain. See Appendix~\ref{app:noise-level}.} $o_\nu$ is the receiver's additive bias (offset or noise temperature), $s_\nu$ is the global 21-cm signal, $f_{\nu,t}$ is the frequency- and time-dependent beam-weighted foreground emission, and $n_{\nu,t}$ is a random realization of a Gaussian noise vector. In order to fit the data, we form the model
\begin{multline}
  \mM_{T_{\nu,t}}(\bzeta_G,\bzeta_f,\bzeta_s,\bzeta_o) =\\ \mM_{G_\nu}(\bzeta_G)\times\left[\mM_{f_{\nu,t}}(\bzeta_f) + \mM_{s_\nu}(\bzeta_s) + \mM_{o_\nu}(\bzeta_o)\right], \label{eq:block-model}
\end{multline}
where $\mM_X(\bzeta_X)$ is the model of the component $X$ and $\bzeta_X$ is the vector of parameters of that model. Figure~\ref{fig:specific-model-tree} shows a tree representation of this model. As described in Section~\ref{sec:pylinex-implementation}, AMLP can be applied to any model that can be represented by this kind of sum-product tree using the \texttt{pylinex} code.

\subsection{Fitting strategy}

The main goal of this fit is a set of constraints on the parameters of a nonlinear signal model. While one may attempt to fit the desired nonlinear signal model directly, the posterior distribution is often difficult to numerically explore if the sampling algorithm is not started in a narrow region around the maximum. Therefore, as is laid out in the following sections, we fit the data with a linear signal model first to home in on a specific set of spectral shapes and follow up with a nonlinear signal model fit initialized near this region.

\subsubsection{Form models} \label{sec:form-models}

  Before we fit the data, we must form models for each of the four components.

  \begin{itemize}
    \item Generate a model to fit the receiver gain $\mM_{G_\nu}(\bzeta_G)$, which can be either linear or nonlinear.
    \item Create linear model of the receiver noise temperature, $\mM_{o_\nu}(\bzeta_o)$, either from a priori knowledge or from a training set of simulated noise temperature curves.\footnote{In this paper, we assume the model of the noise temperature is linear, although modifications to this method for the case of a nonlinear noise temperature model are discussed in Section~\ref{sec:nonlinear-offset-model}.}
    \item Create linear model of beam-weighted foreground spectra, $\mM_{f_{\nu,t}}(\bzeta_f)$, from a training set simulated using reasonably varied beams and foreground maps.
    \item Choose a nonlinear signal model, $\mM_{s_\nu}^{\text{NL}}(\bzeta_s^{\text{NL}})$.
    \item Create linear model, $\mM_{s_\nu}^{\text{L}}(\bzeta_s^{\text{L}})$, of the signal using a training set generated from the chosen nonlinear signal model. \label{item:linear-signal-model}
  \end{itemize}
  
  \subsubsection{Fit using the linear signal model} \label{sec:fit-using-linear-signal-model}
  
  Perform AMLP using the model of Equation~\ref{eq:block-model} with the linear signal model, i.e. $\mM_{s_\nu}(\bzeta_s)=\mM_{s_\nu}^{\text{L}}(\bzeta_s^{\text{L}})$. Split the parameters through $\bthetaNL=\bzeta_G$ and $\bthetaL=\begin{bmatrix} \bzeta_s^{\text{L}} \\ \bzeta_f \\ \bzeta_o \end{bmatrix}$. In this fit, only the parameters of the gain model must be explored numerically, so it should be completed very quickly and will produce a fast estimate of the signal in frequency-temperature space. The walkers of the MCMC should be initialized through a sample of the prior distribution on $\bzeta_G$.
  
  \subsubsection{Prepare to fit using the nonlinear signal model} \label{sec:prepare-to-fit-using-nonlinear-signal-model}
  
  The final fit will need to numerically explore both $\bzeta_G$ and $\bzeta_s^{\text{NL}}$. An estimate of the distribution of $\bzeta_G$ should be available from the fit done in the previous step; but, we must find an estimate of the distribution of $\bzeta_s^{\text{NL}}$.\footnote{Item~\ref{item:linear-signal-model} from Section~\ref{sec:form-models} and the steps laid out in Sections~\ref{sec:fit-using-linear-signal-model}~and~\ref{sec:prepare-to-fit-using-nonlinear-signal-model} can be skipped if the nonlinear signal model is very fast (e.g. if many signals can be evaluated in one second). These steps are meant to provide a starting point for the final fit so that the nonlinear signal model parameter distribution can be achieved in the fewest possible steps.} To do so, we perform a least square fit to the linear signal model parameter mean vector from the first AMLP fit described in Section~\ref{sec:fit-using-linear-signal-model} in mode coefficient space, i.e. we define the first guess nonlinear parameter vector $\bzeta_{s,\text{guess}}^{\text{NL}}$ through
  \begin{equation*}
    \bzeta_{s,\text{guess}}^{\text{NL}}=\underset{\bzeta_{s}^{\text{NL}}}{\text{argmin}}\left|\bLambda_{\text{L}}^{-1/2} \left[\bmM_{s}^{\text{L},-1}(\bmM_{s}^{\text{NL}}(\bzeta_s^{\text{NL}}))-\overline{\bzeta}_s^{\text{L}}\right]\right|^2,
  \end{equation*}
  where $\overline{\bzeta}_s^{\text{L}}=\E[\bzeta_s^{\text{L}}]$, $\bLambda_{\text{L}}=\Cov[\bzeta_s^{\text{L}}]$, and $\bmM_s^{\text{L},-1}$ is the pseudo-inverse of $\bmM_s^{\text{L}}$.\footnote{Here, $\bmM_{s}$ is a vectorized form of $\mM_{s_\nu}$. By pseudo-inverse, we mean the function that takes in a spectrum $\bs$ and outputs the parameters $\bzeta_s^{\text{L}}$ that minimize the difference between $\bs$ and $\bmM_{s}^{\text{L}}(\bzeta_s^{\text{L}})$, as measured by the noise covariance, $\bC$. If $\bmM_{s}^{\text{L}}(\bzeta_s^{\text{L}})=\ba+\bF\bzeta_s^{\text{L}}$, then $\bmM_s^{L,-1}(\bs)=(\bF^T\bC^{-1}\bF)^{-1}\bF^T\bC^{-1}(\bs-\ba)$.} We also derive a covariance of the guess distribution from the Fisher information, i.e.
  \begin{equation*}
    \bLambda_{\text{NL},\text{guess}} = (\bZ^T\bLambda_{\text{L}}^{-1}\bZ)^{-1},
  \end{equation*}
  where $\bZ=\left[\frac{\partial}{\partial\bzeta_s^{\text{NL}}}\bmM_s^{\text{L},-1}(\bmM_s^{\text{NL}}(\bzeta_s^{\text{NL}}))\right]_{\bzeta_{s}^{\text{NL}}=\bzeta_{s,\text{guess}}^{\text{NL}}}$. To initialize walkers for the MCMC of the final nonlinear signal model fit, we draw a sample of gain parameters from the chains of the MCMC fit from Section~\ref{sec:fit-using-linear-signal-model} and we draw the nonlinear signal parameters from the distribution defined by the mean and covariance given in this section, i.e.
  \begin{equation*}
    \bzeta_{s}^{\text{NL}}\sim\mN(\bzeta_{s,\text{guess}}^{\text{NL}},\bLambda_{\text{NL},\text{guess}}).
  \end{equation*}
  
  \subsubsection{Fit using the nonlinear signal model} \label{sec:fit-using-nonlinear-signal-model}
  
  Finally, we perform AMLP using Equation~\ref{eq:block-model} with the nonlinear signal model, i.e. $\mM_{s_\nu}(\bzeta_s)=\mM_{s_\nu}^{\text{NL}}(\bzeta_s^{\text{NL}})$, and splitting the parameters into $\bthetaNL=\begin{bmatrix} \bzeta_G \\ \bzeta_s^{\text{NL}} \end{bmatrix}$ and $\bthetaL=\begin{bmatrix} \bzeta_f \\ \bzeta_o \end{bmatrix}$. Because more parameters are explored numerically, this fit will naturally take longer than the fit with the linear signal model; but, it will ultimately output the desired constraints on the signal parameters.
  
\subsection{Noise level}

The only aspect of the fitting strategy left to define is how to determine the noise covariance $\bC$. As is described in Appendix~\ref{app:noise-level}, the noise in each frequency bin is independent and nearly equal to the data divided by the dynamic range factor $\sqrt{\Delta\nu\ \Delta t}$, where $\Delta\nu$ is the channel width in Hz and $\Delta t$ is the integration time per spectrum in $s$, i.e.
\begin{equation}
  C_{jk} = \frac{(y_j)^2}{\Delta\nu\ \Delta t} \times \begin{cases} 1 & j=k \\ 0 & j\ne k \end{cases}. \label{eq:noise-covariance}
\end{equation}
For the fits in this paper, in addition to being used to define $\bC$, this covariance is used to simulate the random noise $n_{\nu,t}$ from Equation~\ref{eq:observation-equation}. We use $\Delta\nu=1$ MHz and $\Delta t = 800$ hr.

\section{Models} \label{sec:models}

While Section~\ref{sec:fitting-strategy} laid out the general strategy for fitting observations while including receiver models, in this section, we put forth the specific models used in this work, following the tasks from Section~\ref{sec:form-models}.

\begin{table}[t!]
  \centering
  \caption{Gain model parameters} \label{tab:gain-model-parameters}
  \begin{tabular*}{0.47\textwidth}{c @{\extracolsep{\fill}} ccc}
    \hline
    \hline
    Parameter & Unit & Input & Prior \\
    \hline
    $10\log_{10}{\left[1 + \left(\varepsilon^{\text{HP}}\right)^2\right]}$ & dB & $0.25$ & $\text{Unif}(0, 1)$ \\
    $10\log_{10}{\left[1 + \left(\varepsilon^{\text{LP}}\right)^2\right]}$ & dB & $0.25$ & $\text{Unif}(0, 1)$ \\
    $\nu_0^{\text{HP}}$ & MHz & 41.1 & $\text{Unif}(35, 50)$ \\
    $\nu_0^{\text{LP}}$ & MHz & 116.8 & $\text{Unif}(110, 125)$ \\
    \hline
  \end{tabular*}
    \vspace{0.8ex}
    
    {\raggedright \textbf{Notes.} The transformations of the $\varepsilon$ parameters in the first two rows are the levels of the in-band ripples in dB. \par}
\end{table}

\subsection{Gain model} \label{sec:gain-model}

In this paper, we use a gain model that is the product of a low-pass (LP) filter and a high-pass (HP) filter, i.e.
\begin{multline}
  \mM_{G_\nu}\left(\nu_0^{\text{HP}}, \varepsilon^{\text{HP}}, \nu_0^{\text{LP}}, \varepsilon^{\text{LP}}\right) =\\ \mM_{G^{\text{HP}}_\nu}\left(\nu_0^{\text{HP}}, \varepsilon^{\text{HP}}\right)\times \mM_{G^{\text{LP}}_\nu}\left(\nu_0^{\text{LP}}, \varepsilon^{\text{LP}}\right).
\end{multline}
The individual filters are taken to have Chebyshev type I transfer functions, meaning that their gains are given by
\begin{subequations}
\begin{align}
  \mM_{G_\nu^{\text{HP}}}\left(\nu_0^{\text{HP}},\varepsilon^{\text{HP}}\right) &= \frac{1}{1+\left[\varepsilon^{\text{HP}}\ T_n\left(\frac{\nu_0^{\text{HP}}}{\nu}\right)\right]^2}, \\
  \mM_{G_\nu^{\text{LP}}}\left(\nu_0^{\text{LP}},\varepsilon^{\text{LP}}\right) &= \frac{1}{1+\left[\varepsilon^{\text{LP}}\ T_n\left(\frac{\nu}{\nu_0^{\text{LP}}}\right)\right]^2},
\end{align}
\end{subequations}
where $T_n$ is the $n^{\text{th}}$ order Chebyshev polynomial.\footnote{Chebyshev polynomials are defined by $T_n(\cos{\theta})=\cos{n\theta}$. We use $n=9$.} The input gain curve in the simulations (see the red curve in Figure~\ref{fig:nonlinear-uncertainties}) is generated using the parameter values shown in Table~\ref{tab:gain-model-parameters}. The $\varepsilon$ parameters determine the level of ripple inside the band. We choose values that generate a 0.25 dB ripple in the pass-band from both the high-pass and low-pass filters, leading to a total ripple of roughly 0.5 dB. The reference frequencies $\nu_0^{\text{HP}}$ and $\nu_0^{\text{LP}}$ were chosen so that the gain is close to $\frac{1}{2}$ ($-3$ dB) at the edges of the observed band (40-120 MHz).

\subsection{Nonlinear signal model} \label{sec:nonlinear-signal-model}

\begin{figure}[t!]
  \centering
  \includegraphics[width=0.47\textwidth]{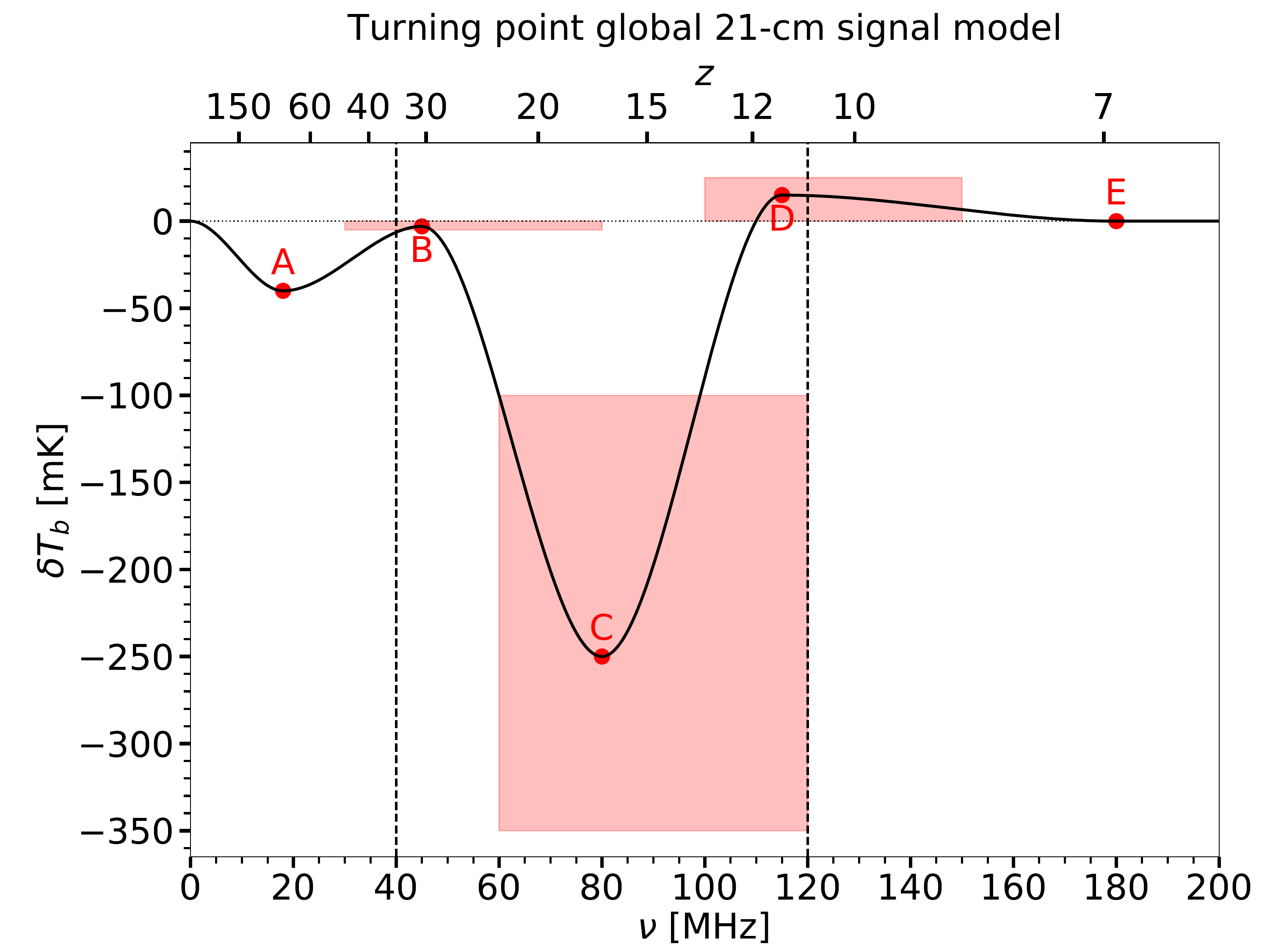}
  \caption{Illustration of the turning point signal model. The signal is a cubic spline using knots at the labeled turning points to enforce that they are local extrema. The red rectangles around turning points B-D indicate the prior distribution of their frequencies and temperatures (see Table~\ref{tab:turning-point-parameters}). Turning points A and E are fixed because they will not be constrained by points within the observed band (40-120 MHz) marked by the vertical dashed lines.} \label{fig:turning-point-explanation}
\end{figure}

Here, we use the so-called turning point model of the signal (used also in Paper II), which is an interpolation between turning points (i.e. local extrema) of the 21-cm signal. These turning points are labeled A-E and are illustrated and described in Figure~\ref{fig:turning-point-explanation}. The free parameters of the model are the frequencies and brightness temperatures of the turning points B-D, whereas $(\nu_A, T_A)$ is fixed to the $\Lambda$CDM value of $(18\text{ MHz}, -40\text{ mK})$ and $(\nu_E,T_E)$ is fixed to $(180\text{ MHz}, 0\text{ mK})$.\footnote{We justify fixing $\nu_E$ by noting that it will not be constrained by observations between 40-120 MHz.} The model is a cubic spline between the turning points. In order to force the turning points to be extrema (i.e. have derivative zero), each turning point uses two spline knots placed at the same temperature and 20 kHz apart symmetrically around the turning point frequency. In addition to turning points A-E, there are two knots placed at 0 K and $10\pm10$ kHz to force the signal to approach zero smoothly at very large redshifts.~The model always evaluates to 0 K at frequencies above that of turning point E. The red rectangles in Figure~\ref{fig:turning-point-explanation} and the last column of Table~\ref{tab:turning-point-parameters} indicate the priors we place on the six varying parameters.

\begin{table}[t!]
    \centering
    \caption{Turning point signal model parameters} \label{tab:turning-point-parameters}
    \begin{tabular*}{0.47\textwidth}{c @{\extracolsep{\fill}} ccc}
        \hline
        \hline
        Parameter & Units & Input & Prior \\
        \hline
        $\nu_A$ & MHz & 18 & Fixed \\
        $T_A$ & mK & -40 & Fixed \\
        $\nu_B$ & MHz & 45 & Unif(30, 80) \\
        $T_B$ & mK & -3 & Unif(-5, 0) \\
        $\nu_C$ & MHz & 80 & Unif(60, 120) \\
        $T_C$ & mK & -250 & Unif(-350, -100) \\
        $\nu_D$ & MHz & 115 & Unif(100, 150) \\
        $T_D$ & mK & 15 & Unif(0, 25) \\
        $\nu_E$ & MHz & 180 & Fixed \\
        \hline
    \end{tabular*}
    \vspace{0.8ex}
    
    {\raggedright \textbf{Notes.} The frequencies of adjacent turning points are also constrained to differ by at least 10 MHz. \par}
\end{table}

\subsection{Linear models} \label{sec:linear-models}

As in Paper I, we form models through decomposition of training sets. For doing so in this paper, though, we apply a slightly modified procedure that employs a principal component analysis-like formalism described in Appendix~\ref{app:pca}. Here, we summarize how we form the training sets in the case of each component. In the following, we denote an individual training set curve of component $X$ as $b_X$.

\begin{itemize}
  \item \textit{Noise temperature training set}: $\mM_{o_\nu}$ is formed from a training set where each curve is a line parameterized by its low- and high-frequency endpoints, $o_L$ and $o_H$, i.e.
  \begin{equation}
    b_{o_\nu}=o_L\left[\frac{(120\text{ MHz})-\nu}{80\text{ MHz}}\right]+o_H\left[\frac{\nu-(40\text{ MHz})}{80\text{ MHz}}\right].
  \end{equation}
  To generate many curves, we assume that $\frac{o_L}{1\text{ K}}\sim\text{Unif}(115, 120)$ and $\frac{o_H}{1\text{ K}}\sim\text{Unif}(110, 115)$.

\begin{figure}[t!]
  \centering
  \includegraphics[width=0.47\textwidth]{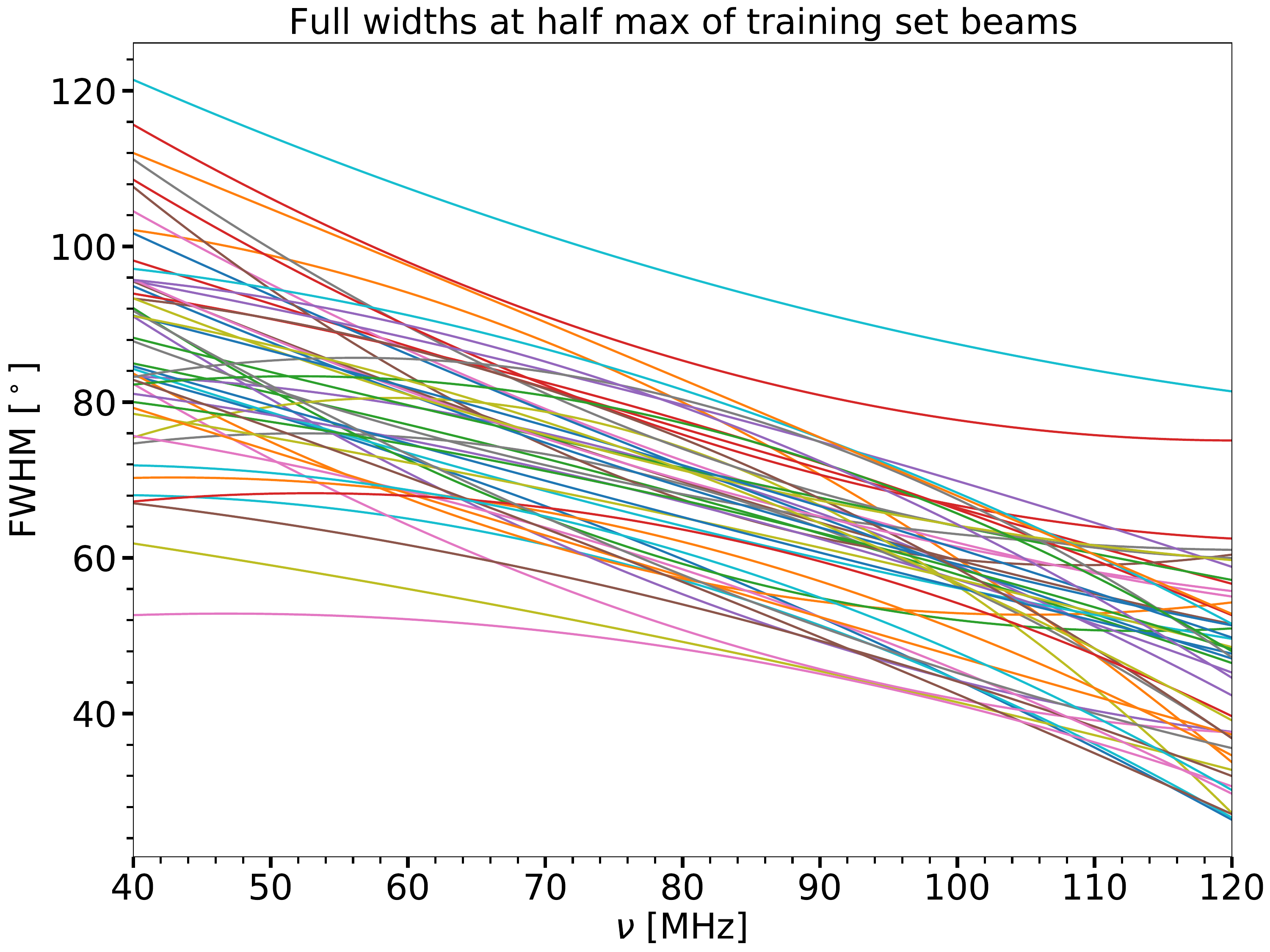}
  \caption{Ten percent of the training set of 500 beam full widths at half maximum (FWHM). The curves are quadratic Legendre polynomials with coefficient distributions given in Table~\ref{tab:legendre-coefficients}.} \label{fig:foreground-training-set}
\end{figure}

\renewcommand{\tabcolsep}{8mm}
\begin{table}[t!]
    \centering
    \caption{Parameters of Legendre coefficient distributions} \label{tab:legendre-coefficients}
    \begin{tabular}{ccc}
        \hline
        \hline
        $k$ & $\mu_k$ & $\sigma_k$ \\
        & $[^\circ]$ & $[^\circ]$ \\
        \hline
        0 & 70 & 10 \\
        1 & -20 & 5 \\
        2 & 0 & 5 \\
        \hline
        \multicolumn{3}{l}{\textbf{Note}: See Equations~\ref{eq:legendre-polynomials},~\ref{eq:legendre-polynomial-definitions},~and~\ref{eq:coefficient-distribution}.}
    \end{tabular}
\end{table}

  \item \textit{Beam-weighted foreground training set}: A general beam-weighted foreground training set should be created from two sources: variations in the antenna beam and in foreground emission. In this paper, as in the rest in the series, however, we use many beams and one foreground map,\footnote{See \cite{Hibbard:20} for an initial attempt at varying the foreground in this formalism.} with the latter given by \cite{Remazeilles:15} \citep[which used the data of][]{Haslam:82} scaled with a spectral index of -2.5.~The beams are angular Gaussians, i.e.~they satisfy $B(\nu,\theta,\phi)\propto \exp{\left\{-\frac{\theta^2}{2[\alpha(\nu)]^2}\right\}}$, where $\theta$ and $\phi$ are the polar and azimuthal spherical coordinate angles, respectively. The scale $\alpha$ is a function of frequency, $\nu$, so that beam chromaticity can be included in the analysis robustly. The full width at half maximum (FWHM), given by $\text{FWHM}(\nu) = \sqrt{8\ln{2}}\ \alpha(\nu)$, is varied between training set curves and is generated by quadratic polynomials in frequency. To control magnitude variations in each order simply, we use second-order Legendre polynomials:
\begin{equation}
  \text{FWHM}(\nu) = \sum_{k=0}^2 a_kL_k\left(\frac{\nu-\nu_0}{\delta\nu}\right), \label{eq:legendre-polynomials}
\end{equation}
where $\nu_0=(\nu_{\text{max}} + \nu_{\text{min}})/2$ is the average frequency, $\delta\nu=(\nu_{\text{max}}-\nu_{\text{min}})/2$ is half the width of the frequency band, and
\begin{equation}
    L_0(x) = 1,\ \ 
    L_1(x) = x,\ \ 
    L_2(x) = \frac{3x^2-1}{2}. \label{eq:legendre-polynomial-definitions}
\end{equation}
In our case, $\nu_{\text{min}}=40$ MHz and $\nu_{\text{max}}=120$ MHz, so $\nu_0=80$ MHz and $\delta\nu=40$ MHz. To seed the beam variations in our training set, we draw $a_0$, $a_1$, and $a_2$ from independent normal distributions,
\begin{equation}
  a_k \sim \mN(\mu_k, \sigma_k^2), \label{eq:coefficient-distribution}
\end{equation}
with the means and standard deviations $\mu_k$ and $\sigma_k$ given in Table~\ref{tab:legendre-coefficients}. An extra constraint is applied to exclude $\text{FWHM}(\nu)$ curves which dip below $15^\circ$ in the $40-120$ MHz band. The resulting training set of $\text{FWHM}$ curves is shown in Figure~\ref{fig:foreground-training-set}.

  We simulate the beam-weighted foreground temperature for ten different bins in local sidereal time (LST) by smearing the map through LST before computing the spectra.\footnote{By smearing, we refer to averaging many rotations of the foreground map corresponding to LSTs between the edges of each LST bin.}~The spectra from the ten LST bins are then concatenated into the final data vector.
  \item \textit{Signal training set}: $\mM^{\text{L}}_{s_\nu}$ is formed from a training set where each curve is created by applying the turning point nonlinear signal model $\mM^{\text{NL}}_{s_\nu}$ to a sample from the prior distribution of its parameters (see Section~\ref{sec:nonlinear-signal-model}, Figure~\ref{fig:turning-point-explanation}, and Table~\ref{tab:turning-point-parameters}).
\end{itemize}


\section{Results} \label{sec:results}

This section lays out the results of applying the general AMLP method described in Section~\ref{sec:AMLP} to the procedure put forth in Section~\ref{sec:fitting-strategy} for fitting sky-averaged radio data measured with a non-ideal receiver using the models laid out in Section~\ref{sec:models}. The initial fit using the linear signal model (following Section~\ref{sec:fit-using-linear-signal-model}) is shown in Section~\ref{sec:results-fit-using-linear-signal-model} and the final fit using the nonlinear signal model (following Section~\ref{sec:fit-using-nonlinear-signal-model}) is shown in Section~\ref{sec:results-fit-using-nonlinear-signal-model}.

\begin{figure}[t!]
  \centering
  \includegraphics[width=0.43\textwidth]{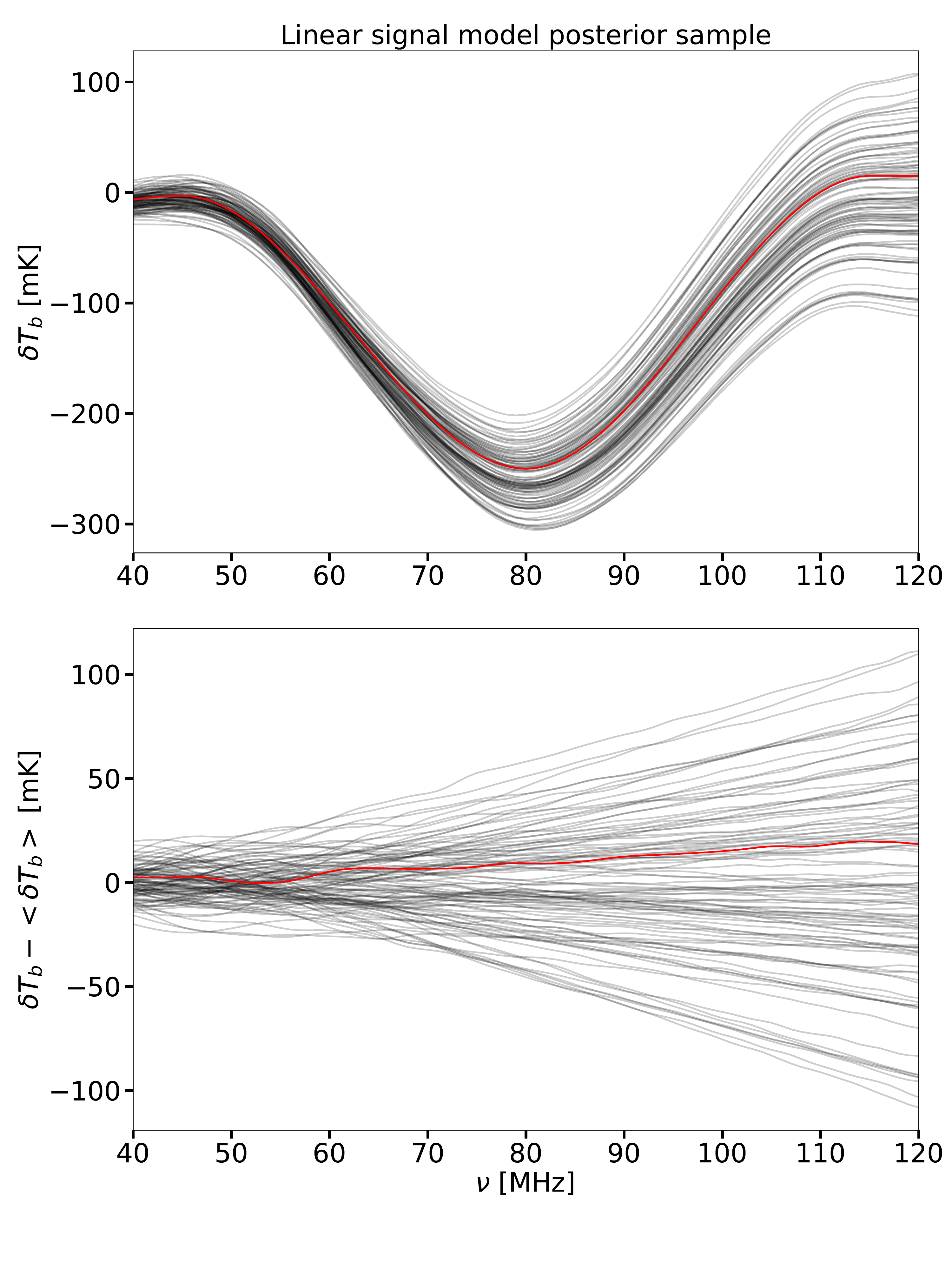}
  \caption{\textit{Top}: Signal realizations from the first MCMC fit with the linear signal model, i.e. $\mM_{s_\nu}^{\text{L}}(\bzeta_s^{\text{L}})$, where $\bzeta_s^{\text{L}}$ is sampled from the posterior distribution, are shown in black. The red line shows the input signal, which is clearly inside the uncertainty interval. \textit{Bottom}: Same as top with the mean of the black lines subtracted from each curve. The main source of variance in the signal estimate comes from modes similar to lines, which are nearly degenerate with the linear modes describing the noise temperature.} \label{fig:linear-signal-model-uncertainties}
\end{figure}

\subsection{Fit with linear signal model} \label{sec:results-fit-using-linear-signal-model}

The signal uncertainties implied by the first MCMC fit with the linear signal model are shown in Figure~\ref{fig:linear-signal-model-uncertainties}. The main mode of uncertainty is line-like, with a width of approximately 50 mK. This width comes from the similarity, or overlap, between the linear signal model and the models of the other components. For simplicity, here we will discuss the effect of a similarity between the signal and noise temperature models, but it is important to note that similarities between the signal model and the gain or beam-weighted foreground models would produce the same effect. The uncertainties implied by the posterior are designed to answer a fundamental question: what size shift in signal parameters $\delta\bzeta_s^{\text{L}}$ can lead to a spectral change that is compatible with the Gaussian noise distribution of the data when accounting for offsetting changes $\delta\bzeta_o$ in parameters of the noise temperature model? If all combinations of $\delta\bzeta_s^{\text{L}}$ and $\delta\bzeta_o$ produce orthogonal effects on the data, then the allowed size of $\delta\bzeta_s^{\text{L}}$ is determined by the noise alone, and the resulting signal uncertainties should essentially match the noise level of the data. On the other hand, if the effects of some $\delta\bzeta_s^{\text{L}}$ can be offset very closely by a corresponding change $\delta\bzeta_o^{\text{NL}}$, then the uncertainties will be larger to account for this overlap. Moreover, if both the signal and noise temperature models are linear (as they are in this first MCMC fit) and they have overlapping gradients, then the uncertainties can grow greatly or even diverge because if, for example, $\delta\bzeta_s^{\text{L}}$ can be exactly offset by $\delta\bzeta_o$, then $2\delta\bzeta_s^{\text{L}}$ can be exactly offset by $2\delta\bzeta_o$. The uncertainties shown in the bottom panel of Figure~\ref{fig:linear-signal-model-uncertainties} are caused by the fact that there is a direction in the $\bzeta_s^{\text{L}}$ parameter space that causes a line-like spectral feature. Since the noise temperature model used in this paper is a linear function of frequency, it can closely offset a line-like signal change, leading to large signal errors with a line-like shape.

The signals from this first MCMC fit are used to initialize the MCMC fit with the nonlinear signal model as described in Sections~\ref{sec:prepare-to-fit-using-nonlinear-signal-model}~and~\ref{sec:fit-using-nonlinear-signal-model}.

\begin{figure}[t!]
  \centering
  \includegraphics[width=0.47\textwidth]{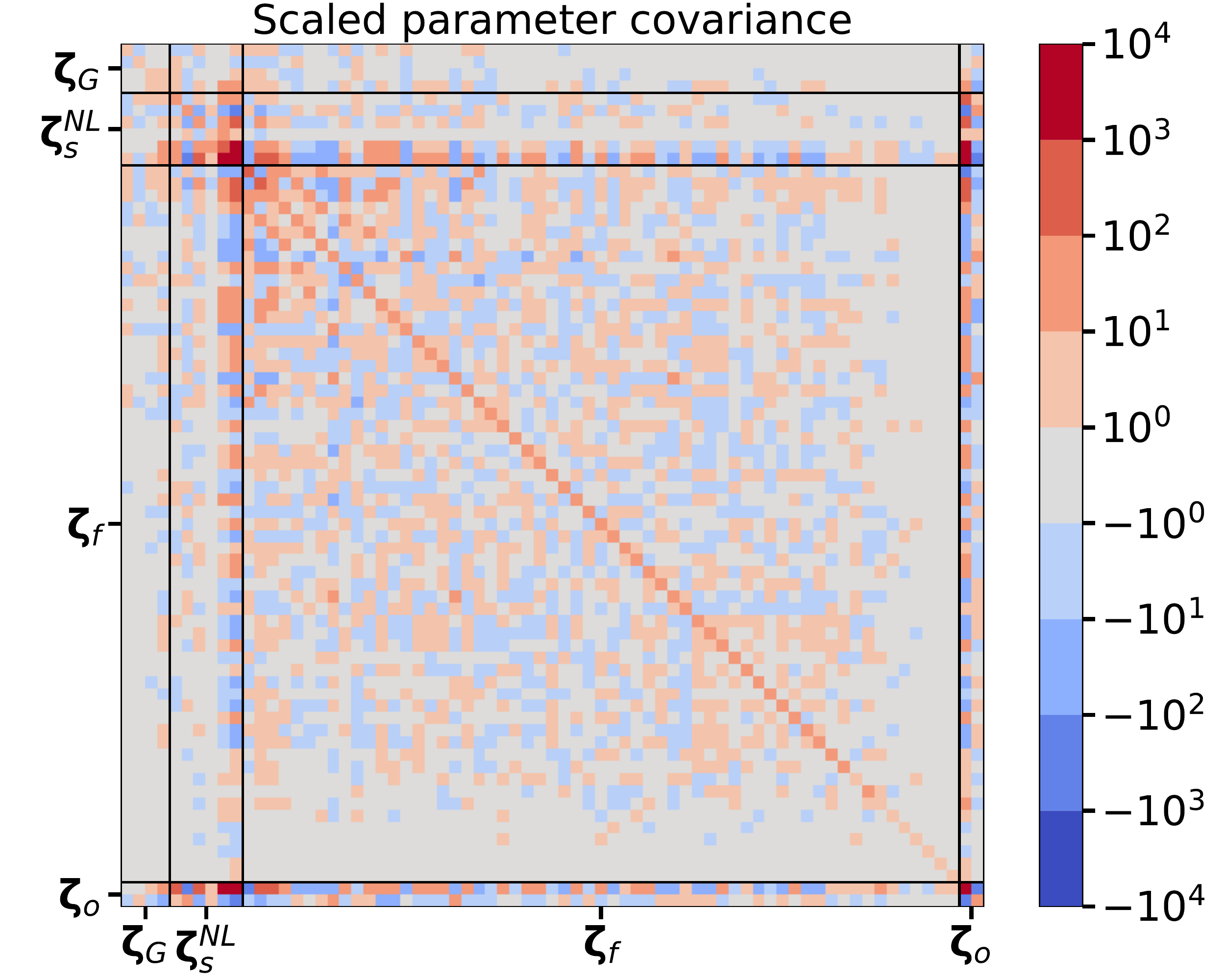}
  \caption{The full parameter covariance matrix arising from the final posterior distribution in the fit with the nonlinear signal model.~Each row and column is normalized so that diagonal elements of one correspond to a difference of one noise level in a sum of squares sense.~The blocks of the matrix corresponding to the parameters of different components (i.e. gain, signal, foreground, and noise temperature) are separated by black lines. The noise temperature is the component that covaries most with the signal because it appears similarly in the observation equation (Equation~\ref{eq:observation-equation}).} \label{fig:nonlinear-fit-parameter-covariance}
\end{figure}

\begin{figure*}[h!]
  \centering
  \includegraphics[width=0.423\textwidth]{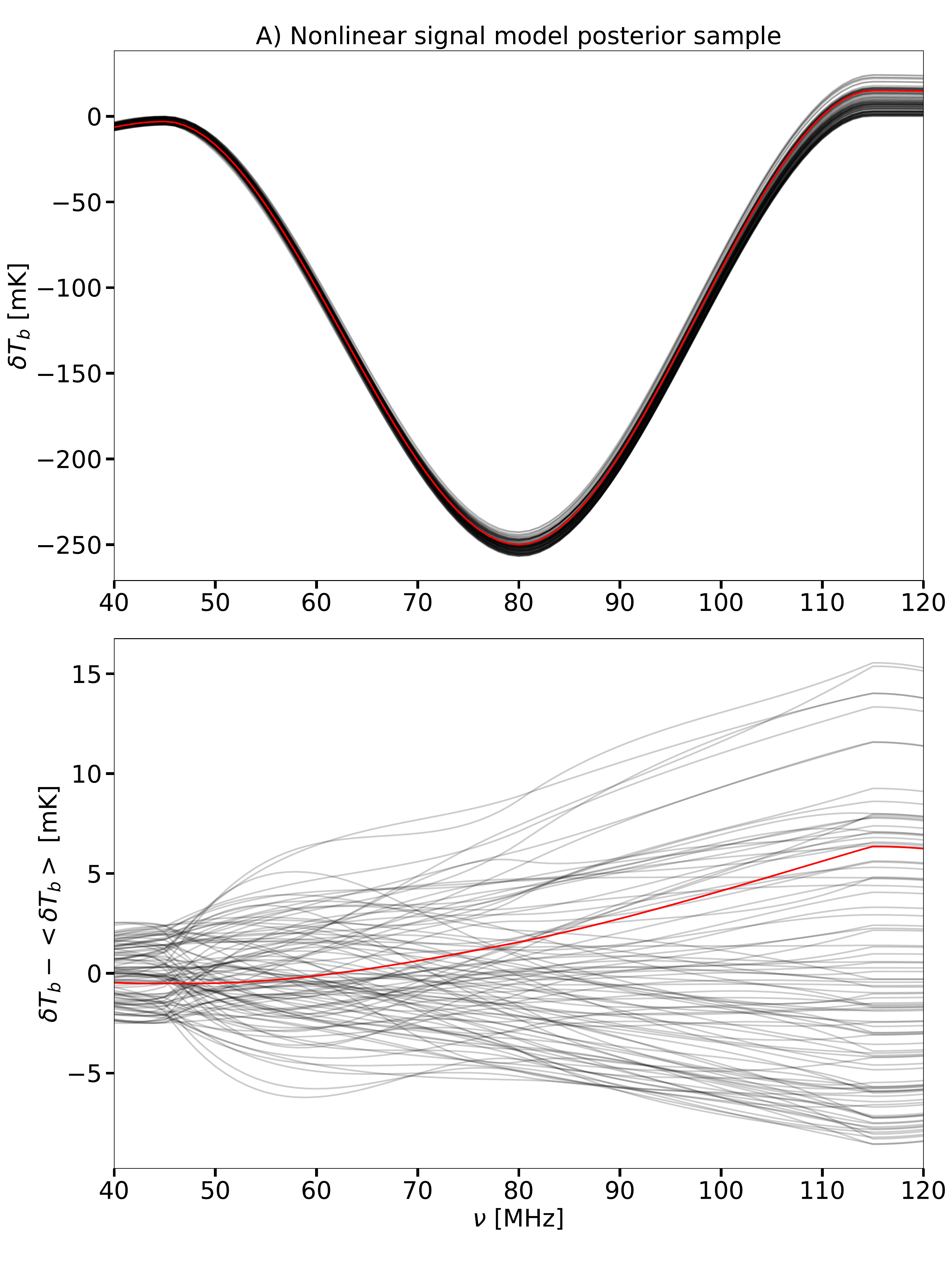}
  \hspace{1cm}
  \includegraphics[width=0.423\textwidth]{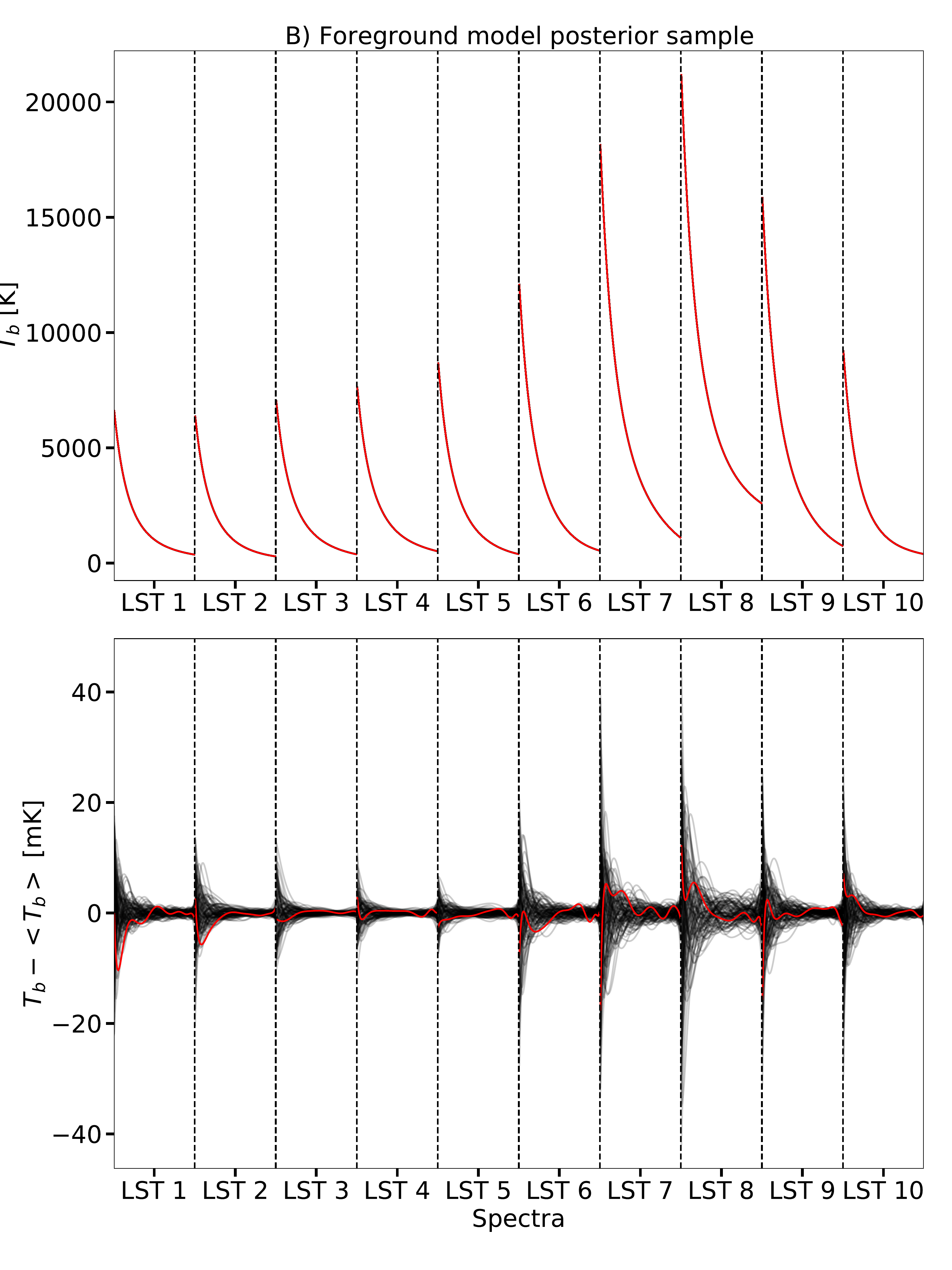}
  \includegraphics[width=0.423\textwidth]{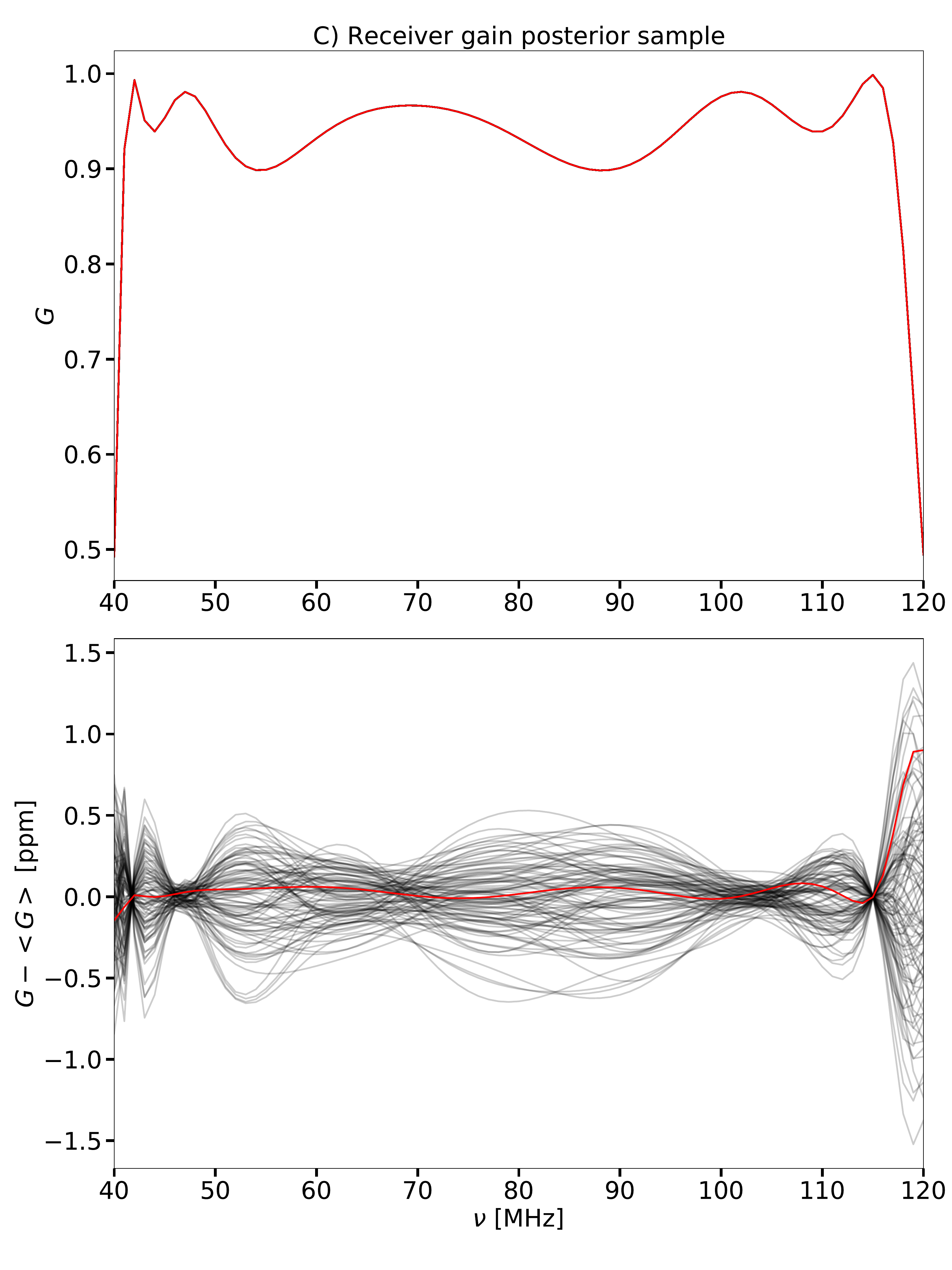}
  \hspace{1cm}
  \includegraphics[width=0.423\textwidth]{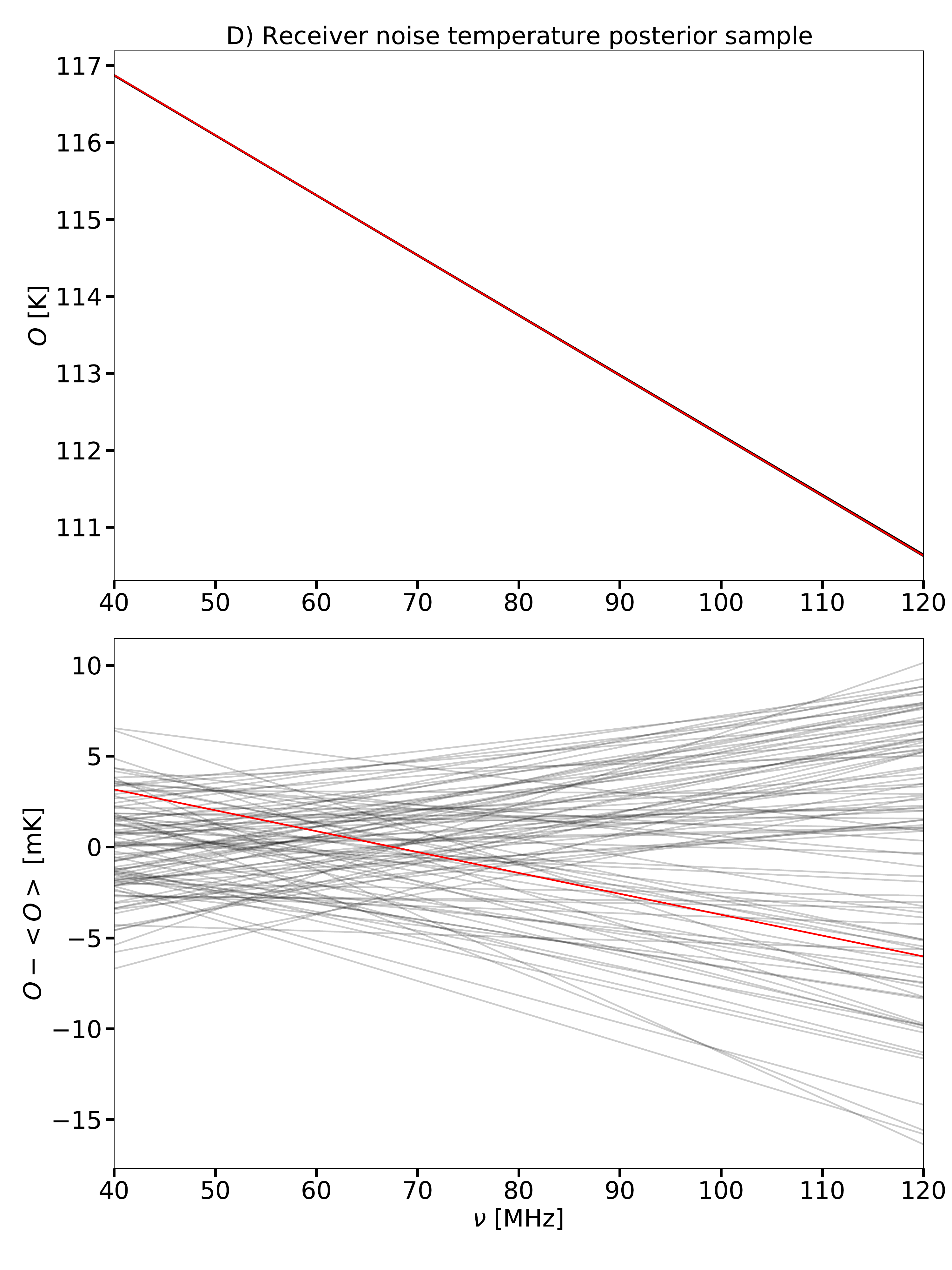}
  \caption{Each plot shows 100 curves (black lines) from the final MCMC-explored posterior distribution with the nonlinear signal model alongside the input (red line) for the given component. The lower of each pair of plots shows the uncertainties on that component by subtracting the mean of the black curves from the plot above it. The two plots in the top left corner (A) show the signal uncertainties, which are much improved from their counterparts in Figure~\ref{fig:linear-signal-model-uncertainties}. The beam-weighted foreground sample is shown in the top right pair of plots (B), where the 10 spectra included in the data vector are concatenated. The receiver gain and noise temperature samples are shown in the bottom left (C) and bottom right (D) plots, respectively.} \label{fig:nonlinear-uncertainties}
\end{figure*}

\begin{figure*}[t!]
  \centering
  \includegraphics[width=0.95\textwidth]{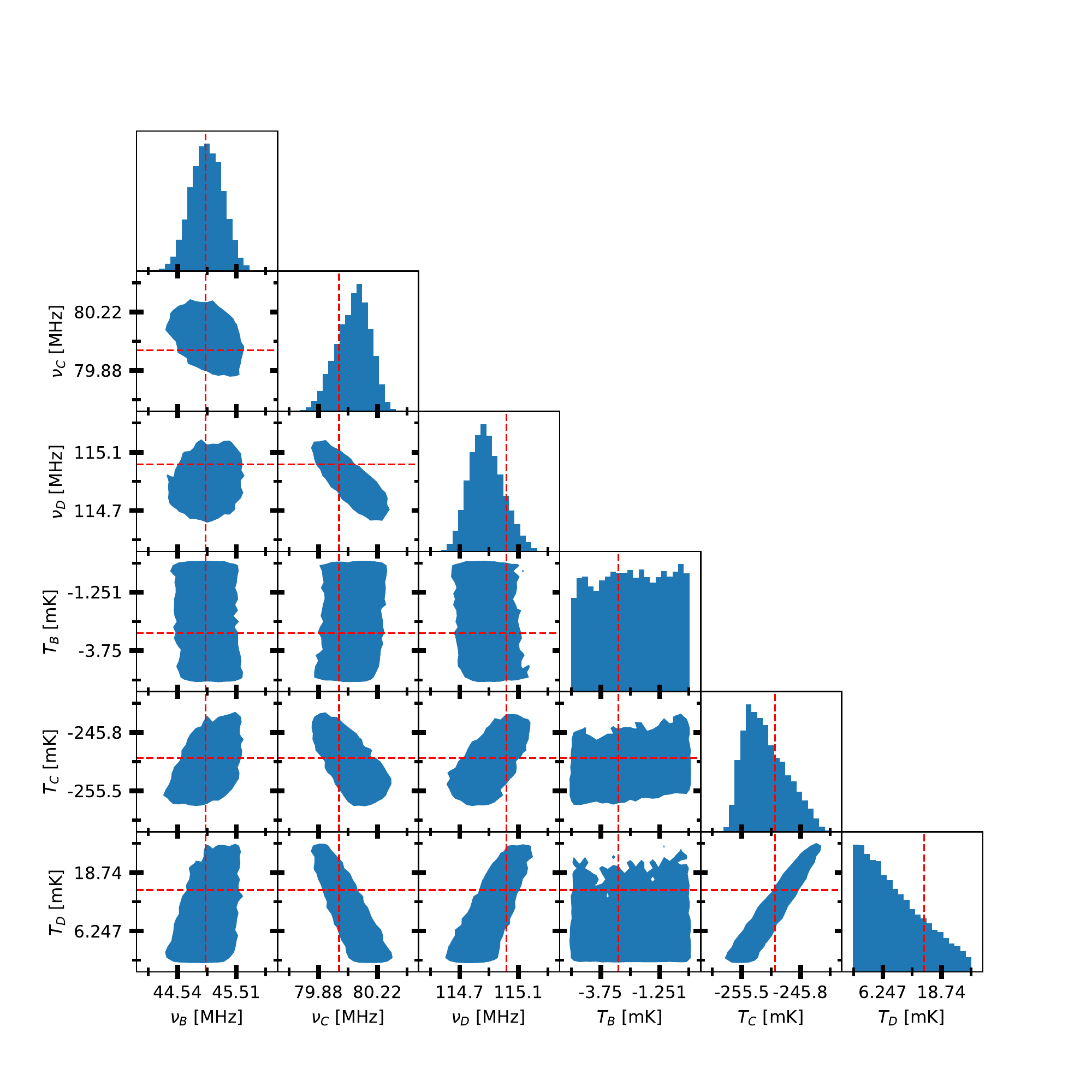}
  \caption{Triangle plot showing marginal distributions on the turning point signal parameters. The univariate marginal distributions of each parameter are shown on the diagonal, while the 95\% confidence intervals of the bivariate marginal distributions are shown in the off-diagonal plots. The dashed lines show the input parameters, which are inside the confidence intervals in each plot. While the temperatures of turning points B and D are not constrained (in a univariate sense) at the 95\% level beyond the narrow priors, the other parameters are tightly constrained.} \label{fig:triangle-plot}
\end{figure*}

\subsection{Fit with nonlinear signal model} \label{sec:results-fit-using-nonlinear-signal-model}

Figure~\ref{fig:nonlinear-fit-parameter-covariance} shows the posterior parameter covariance of the MCMC fit with the nonlinear signal model.~Each row and column is scaled in such a way that a value of unity in the diagonal elements corresponds to a variation of one noise level. The signal covaries most with the noise temperature because the two components appear similarly in the data, i.e. the same in each spectrum, and must be separated based on their spectral shape alone.

The final MCMC fit yields our best mean and uncertainty estimates of the signal, foreground, receiver gain, and receiver noise temperature, which are shown in Figure~\ref{fig:nonlinear-uncertainties}. The signal sample shown in Figure~\ref{fig:nonlinear-uncertainties}A is greatly improved over the signal sample found using the linear signal model (Figure~\ref{fig:linear-signal-model-uncertainties}) for multiple reasons:
\begin{enumerate}
  \item The turning point signal model has fewer freely varying parameters (6) than the linear signal model (33).
  \item Due to it being nonlinear, the gradient of the turning point signal model is not constant, meaning that a given variation $\delta\bzeta_s^{\text{NL}}$ in the parameter vector affects the signal differently based on the point $\bzeta_s^{\text{NL}}$ around which the parameters are varying. Using the terminology of Section~\ref{sec:results-fit-using-linear-signal-model}, this means that even if a signal parameter change $\delta\bzeta_s^{\text{NL}}$ can be exactly offset by a noise temperature parameter change of $\delta\bzeta_o$, it is generally not true that a signal parameter change of $2\delta\bzeta_s^{\text{NL}}$ can be offset by a noise temperature parameter change of $2\delta\bzeta_o$.
\end{enumerate}

Figure~\ref{fig:nonlinear-uncertainties}B shows the foreground sample from the final MCMC fit, with all of the spectra concatenated. The uncertainties encompass the input foreground, shown with the red line, and have widths of 20-60 mK at low frequencies and a few mK at high frequencies. The beam-weighted foreground must be fit to this level to extract the signal without bias.\footnote{In this section, we use ``bias'' to refer to a disparity between a) the difference between the input curve and fitted curve of a particular component and b) the posterior uncertainty band of that component. This means that fits where the mean of the posterior sample is different than the input curve, which is almost always the case, can be unbiased as long as the uncertainties accurately characterize the difference.}

Figure~\ref{fig:nonlinear-uncertainties}C~and~\ref{fig:nonlinear-uncertainties}D show the posterior samples of the receiver gain and noise temperature, respectively. The gain is fit down to the sub-parts per million (ppm) level due to the fact that the dynamic range of the measurement is $\frac{\sqrt{C_{kk}}}{y_k}=\frac{1}{\sqrt{\Delta\nu\ \Delta t}}\approx 0.6$ ppm (see Equation~\ref{eq:noise-covariance} and Appendix~\ref{app:noise-level}) and the noise temperature is fit down to the 10 mK level, which is necessary to fit the signal down to the same level.

A triangle plot showing the univariate and bivariate marginal distributions of the turning point signal parameters is shown in Figure~\ref{fig:triangle-plot}. The temperatures of turning points B and D are not constrained tighter than their narrow priors (see Table~\ref{tab:turning-point-parameters} and Figure~\ref{fig:turning-point-explanation}), although $T_D$ has interesting nontrivial correlations with the frequency and temperature of turning point C. The other four parameters are well constrained. The input values of the parameters are within the posterior uncertainties.

\section{Discussion} \label{sec:discussion}

In this section, we discuss complexities beyond those accounted for in the fits in Section~\ref{sec:results} for the sake of simplicity and how the method can be modified to deal with them in future work.

\subsection{Receiver time-dependence} \label{sec:receiver-time-dependence}

In this paper, we have assumed that the noise temperature and gain are functions of frequency; but, in reality, both the gain and noise temperature of the receiver will also vary with time.

The main source of variation in the receiver is the ambient temperature of the components in the analog radio-frequency signal chain.\footnote{Both gain and noise temperature are functions of the input antenna impedance, which can vary with ambient temperature due to thermal expansion of the antenna components. However, for most cases, this change is relatively negligible comparing to the actual thermal-coupled variations of the electronic components.}~To a large extent, the effects of ambient temperature on the gain and noise temperature can be characterized well in the laboratory. For example, both may rise uniformly in frequency with increasing temperature. In this case, the time dependence of both the gain and noise temperature can be encapsulated in a simple temperature parameter for each LST bin, which would be explored via MCMC when applying the techniques demonstrated in this work.\footnote{As the ambient temperature can be measured alongside the temperatures in the field and saved as metadata, these temperatures would have tight priors and should therefore be sampled efficiently by the MCMC algorithm.}

A common method of calibrating variations due to temperature drift is to model the receiver's frequency response within a certain operating temperature range. A full receiver circuit model can be carefully constructed and constrained by laboratory measurements to determine the instrument state when operating at different temperatures.\footnote{The gain and the noise temperature of a two-port network can be quantified by two sets of network parameters, namely the complex scattering $S$-parameters and noise parameters, which are constrained by fitting circuit models to laboratory measurements, similarly to the technique that was demonstrated in \cite{Nhan:19}.} In addition, the instrument can have internal sources, such as broad-band noise sources, that aid in calibration.~However, even after calibration, uncertainty in the model must be accounted for in order to produce rigorous constraints.~Therefore, in this case, we must form a model for $G_{\nu,t}/\overline{G}_{\nu,t}$ where $\overline{G}_{\nu,t}$ is the gain curve assumed for calibration.~Similarly a model for $O_{\nu,t}-\overline{O}_{\nu,t}$ must be formed to allow for noise temperature variations in a fit to calibrated data.

While the receiver gain was considered to be a deterministic quantity in the fits presented in this paper, there are also stochastic perturbations that generally introduce $1/f$ noise which in turn creates a noise floor, below which the data cannot be averaged down regardless of integration time. In order to reach the ppm-level dynamic range necessary to extract the global signal, the instrument must be dynamically stabilized through state-dependent corrections so that the noise floor is not reached until after 1000 hours of integration.

\subsection{Pass-band ripples}

While the gain and noise temperature models given in Section~\ref{sec:models} are good approximations to the gain and noise temperature when the receiver is designed to fit them, the in-band ripples will not exactly match any analytical model such as that given by the type-I Chebyshev filter transfer function, due to circuit components that vary from their nominal parameters within some manufacturing tolerance and exhibit parasitic behavior and impedance mismatches between the antenna and receiver. The gain model will need to be modified in order to prevent a modeling error from biasing the fitted signal. One way of including these non-ideal ripples is by forming both the gain and noise temperature models via training sets as performed in Section~\ref{sec:linear-models} instead of as an a priori known function like the Chebyshev filter gain model. To create such training sets, the same receiver circuit model mentioned in Section~\ref{sec:receiver-time-dependence} can be evaluated at a large number of physically reasonable variations in the characteristics of individual electronic components. CTP and DAPPER will adopt a similar scheme to create their receiver gain and noise temperature models.\footnote{The CTP and DAPPER receivers will be designed using a type-II (inverse) Chebyshev filter gain model, which has flatter spectral behavior in the pass-band.}

\subsection{Correlations between gain and noise temperature} \label{sec:nonlinear-offset-model}

In this paper, we have implicitly assumed that variations in the gain are independent from variations in the noise temperature; but, this is not true in reality. For instance, as the physical temperature of the receiver increases, the noise temperature and gain should both vary across the band.~Taking advantage of these correlations in the fit should improve it because it would force the model to assign a smaller probability to variations in the gain and noise temperature that do not correspond to each other. To do this, we could enforce that the gain and noise temperature models share some parameters.~For instance, the gain model and noise temperature model could be written $\mM_{g_\nu}(\bzeta_{G+o},\bzeta_G)$ and $\mM_{o_\nu}(\bzeta_{G+o},\bzeta_o)$, respectively, where $\bzeta_G$ ($\bzeta_o$) represents the parameters that are only relevant to gain (noise temperature) variations and $\bzeta_{G+o}$ represents parameters (such as temperature) that cause correlated variations in the gain and noise temperature. For both uses of AMLP in this paper, we could then numerically explore $\bzeta_G$, $\bzeta_o$, and $\bzeta_{G+o}$ instead of marginalizing over any of them since the majority of the benefit of AMLP comes from marginalizing over beam-weighted foreground parameters. 

\subsection{Polarization measurements}

In addition to laying out that using multiple correlated spectra (as done in this paper) vastly improves global signal fits, Paper III showed that posterior signal uncertainties can be significantly improved by including full-Stokes polarization measurements, which were not included in this paper for simplicity. To fit them, the observation equation (\ref{eq:observation-equation}) must be modified. For example, in the case of a dual-dipole antenna, it becomes
\begin{widetext}
\begin{equation}
  \underbrace{\begin{bmatrix} I \\ Q \\ U \\ V \end{bmatrix}}_{T} = \underbrace{\begin{bmatrix} \frac{1}{2}(|g_X|^2+|g_Y|^2) & \frac{1}{2}(|g_X|^2-|g_Y|^2) & 0 & 0 \\ \frac{1}{2}(|g_X|^2-|g_Y|^2) & \frac{1}{2}(|g_X|^2+|g_Y|^2) & 0 & 0 \\ 0 & 0 & \real(g_X^\ast g_Y) & -\imag(g_X^\ast g_Y) \\ 0 & 0 & \imag(g_X^\ast g_Y) & \real(g_X^\ast g_Y) \end{bmatrix}}_{G} \left( \underbrace{\begin{bmatrix} f_I \\ f_Q \\ f_U \\ f_V \end{bmatrix}}_{f} + \underbrace{\begin{bmatrix} s \\ 0 \\ 0 \\ 0 \end{bmatrix}}_{s} + \underbrace{\begin{bmatrix} o_X+o_Y \\ o_X-o_Y \\ 0 \\ 0 \end{bmatrix}}_{o} \right) + \underbrace{\begin{bmatrix} n_I \\ n_Q \\ n_U \\ n_V \end{bmatrix}}_{n},
\end{equation}
\end{widetext}
where $I$, $Q$, $U$, and $V$ are the measured Stokes parameters to be fit, $o_Z$ and $g_Z$ are the noise temperature and complex voltage gain of the $Z$ dipole feed (where $Z$ is either $X$ or $Y$), $f_P$ and $n_P$ are the beam-weighted foreground and Gaussian noise components in the Stokes parameter $P$, and the $\nu$ and $t$ dependence of all quantities involved has been left off for clarity. As shown by the brackets beneath the individual terms, this is Equation~\ref{eq:observation-equation} with $T$, $f$, $s$, $o$, and $n$ generalized to be vectors and $G$ generalized to be a matrix.~It is important to note that even under this generalization, $T$ is still linear in $f$, $s$, and $o$ at constant $G$; so, the methods of this paper still apply.

CTP and DAPPER are planned to have a four-channel correlation receiver that treats the $+X$, $-X$, $+Y$, and $-Y$ antennas as monopoles. This can be described by another straightforward generalization of the observation equation with the Stokes parameters replaced by the auto- and cross-correlations of the monopole voltage signals, each with its own complex voltage gain and noise. Under this formalism, the Stokes parameters can be formed from cross-correlations alone, meaning that the uncorrelated noise of the monopoles can be avoided. Sources of noise that are correlated between different channels of the receiver will still produce Stokes parameter profiles. But, the vast majority of the noise should be uncorrelated between the channels; so, this four channel correlation receiver can avoid most additive biases.

\section{Conclusions} \label{sec:conclusions}

This paper concludes the series on our data analysis pipeline for global 21-cm signal experiments, which was designed as a rigorous alternative to the most common methods used by existing global signal experiments, such as the use of polynomial models.~We have introduced several key ideas throughout the series:
\begin{itemize}
  \item The beam-weighted foreground must be fit down to the noise level by the chosen model. The best way to achieve this is to build a model specific to the given experimental situation (e.g. antenna design, pointing direction, location, etc.).~We do so through matrix decompositions of training sets such as singular value decomposition or principal component analysis.
  \item Signal parameter constraints can be explored rigorously and efficiently by analytically marginalizing over as many parameters as possible while numerically sampling only the essentially nonlinear parameters, i.e. those that leave the model linear if they are fixed.
  \item Uncertainties are vastly improved when including more than one spectrum of data when fitting. For example, modeling the correlations between foreground spectra at different times can decrease uncertainties from the $\sim1$ K level to the $\sim10$ mK level, required for detection of the expected signal, for total integration times in the hundreds of hours.~Including measurements of all four Stokes parameters decreases the uncertainties down nearer to the few mK noise level, allowing for precision cosmology.
  \item While the receiver systematic effects introduce nonlinearity to the model and new potential for confusion when extracting the global signal, they can be rigorously included in the pipeline using the AMLP methodology described in this work. When is well characterized via lab measurements and simulated models, the receiver does not significantly impact the precision of signal extraction.
\end{itemize}

In addition to the points above focused on the global signal, the analytical marginalization of linear parameters (AMLP) technique introduced in this paper can be used to explore any posterior distribution that has Gaussian conditional distributions more efficiently. AMLP is also included in \texttt{pylinex},\footnote{Download available at \url{https://bitbucket.org/ktausch/pylinex}.} the general, publicly available Python code that implements the fitting procedures laid out throughout the series \citep{pylinex:21}.

The pipeline is built into the design of the newest version of the Cosmic Twilight Polarimeter (CTP) and the proposed Dark Ages Polarimeter PathfindER (DAPPER), allowing for training sets to be developed through lab measurements (receiver), theory (signal), external observations (foreground), and simulations (antenna beam and receiver) before observations begin.

\acknowledgments{\noindent We thank Eric Switzer for the early suggestion of numerically exploring only signal parameters, an idea which has evolved into AMLP. D.R.~was supported in the early stages of this work by a NASA Postdoctoral Program Senior Fellowship at the NASA Ames Research Center, administered by the Universities Space Research Association under contract with the National Aeronautics and Space Administration (NASA). This work was also supported by NASA under award number NNA16BD14C for NASA Academic Mission Services.~B.N.~is a Jansky Fellow of the National Radio Astronomy Observatory.~This work is directly supported by the NASA Solar System Exploration Virtual Institute cooperative agreement 80ARC017M0006.}

\software{numpy, scipy, matplotlib, distpy+pylinex}

\newpage

\bibliographystyle{aasjournal}
\bibliography{references}

\appendix

\section{Noise distribution} \label{app:noise-level}

The voltage distribution of frequency samples from the antenna are independent zero-mean circularly symmetric complex Gaussian random variates, $V_{\nu,t} \sim \mCN(0,f_{\nu,t}+s_\nu)$.\footnote{This follows from the discrete Fourier transform matrix being unitary and the voltage time samples being independent zero-mean Gaussian random variables. Note also that we are using a convention that implies that the voltages are given in units of $\sqrt{K}$.} After going through the receiver but before going through the square-law detector, the voltages are $g_\nu(V_\nu + V_{o_\nu})\sim \mCN\big(0, G_\nu(f_{\nu,t}+s_\nu+o_\nu)\big)$, where $g_\nu$ is the complex receiver gain at frequency $\nu$ and $V_{o_\nu}$ is the receiver noise voltage at frequency $\nu$, which satisfies $V_{o_{\nu}}\sim\mCN(0,o_\nu)$. The power in that frequency bin, $P_{\nu,t}=|g_\nu(V_{\nu,t}+V_{o_\nu})|^2$, is therefore Gamma-distributed, $P_{\nu,t}\sim \Gamma(1,T_{\nu,t})$ where $T_{\nu,t}=G_\nu(f_{\nu,t}+s_\nu+o_\nu)$.\footnote{Note that this is the same as Equation~\ref{eq:observation-equation} without the noise term, as this is the deterministic component of the data.} Assuming that $g_{\nu}$ is constant in time (see Section~\ref{sec:receiver-time-dependence}), if $N$ spectra of $P_{\nu,t}$ are combined into an average $\overline{P}_{\nu,t}$, then that average spectrum is also Gamma-distributed, $\overline{P}_{\nu,t}\sim\Gamma\left(N,\frac{1}{N}T_{\nu,t}\right)$. This implies that $\E[\overline{P}_{\nu,t}]=T_{\nu,t}$ and $\sqrt{\Var[\overline{P}_{\nu,t}]} = T_{\nu,t}/\sqrt{N}$. As $N$ grows large, the distribution of $\overline{P}_{\nu,t}$ approaches a Gaussian distribution with this mean and standard deviation. For Nyquist sampling, a raw spectrum with resolution $\Delta\nu$ takes a time equal to $1/\Delta\nu$. Therefore, the number of spectra $N$ is equal to $\Delta\nu\ \Delta t$, where $\Delta t$ is the time spent averaging per spectrum. Thus, for large integration times, $\Delta t$,
\begin{equation}
  \overline{P}_{\nu,t} \overset{\Delta t\rightarrow\infty}{\sim} \mN\left(T_{\nu,t},\frac{T_{\nu,t}^2}{\Delta\nu\ \Delta t}\right).
\end{equation}
Since the observed data points are realizations of $\overline{P}_{\nu,t}$ and we know that $\overline{P}_{\nu,t}$ is very close to $T_{\nu,t}$ for sufficiently large integration times, we can conclude that the $1\sigma$ noise level of the data is the data itself divided by $\sqrt{\Delta\nu\ \Delta t}$.

\section{Making affine models from training sets} \label{app:pca}

For the purpose of this section, we assume there is a training set matrix $\bB$ that has $n_c$ rows (channels) and $n_t$ columns (training set examples). We wish to find a fixed vector $\balpha$ and a fixed matrix $\bF$ that has $n_c$ rows and $n_b$ columns (basis vectors) such that the model $\boldm(\bzeta)=\balpha+\bF\bzeta$ best fits the training set, with respect to a noise level given by the positive definite matrix $\bC$. For a given column $\bb$ of $\bB$ (i.e. a given training set curve), the value of $\bzeta$ that minimizes the chi-squared statistic $\chi^2(\bzeta)=[\bb-\boldm(\bzeta)]^T\bC^{-1}[\bb-\boldm(\bzeta)]$ is given by
\begin{equation}
  \bzeta_{\text{opt}} = (\bF^T\bC^{-1}\bF)^{-1}\bF^T\bC^{-1}(\bb-\balpha). \label{eq:zeta-opt}
\end{equation}
The chi-squared statistic evaluated at this parameter vector, $\chi^2_{\text{min}}=\chi^2(\bzeta_{\text{opt}})$, is
\begin{equation}
  \chi^2_{\text{min}} = (\bb-\balpha)^T\bC^{-1}\bPhi(\bb-\balpha),
\end{equation}
where $\bPhi=\bI-\bF(\bF^T\bC^{-1}\bF)^{-1}\bF^T\bC^{-1}$ is the matrix that projects out the column space of $\bF$. Performing this for every training set curve and summing them up yields the total chi-squared statistic,
\begin{equation}
  \chi^2_{\text{total}}(\balpha,\bF) = \Tr[(\bB-\balpha\bj^T)^T\bC^{-1}\bPhi(\bB-\balpha\bj^T)],
\end{equation}
where $\Tr$ denotes the trace operation and $\bj$ is a column vector of $n_t$ ones. Minimizing this subject to the normalization condition $\bF^T\bC^{-1}\bF=\bI$ leads to $\balpha=\frac{1}{n_t}\bB\bj$\footnote{This $\balpha$ is the average of the columns of $\bB$, i.e. the average of all training set curves.} and the columns of $\bF$ being $\bC^{1/2}$ times the first $n_b$ eigenvectors of\footnote{Here, we assume that the eigenvectors are ordered from highest to lowest eigenvalue. Note that the eigenvectors described here are orthonormal (and thus lead to a matrix $\bF$ satisfying our normalization condition) because $\bS$ is a symmetric matrix.}
\begin{equation}
  \bS = \frac{1}{n_t}\bC^{-1/2}\bB\left(\bI-\frac{\bj\bj^T}{n_t}\right)\bB^T\bC^{-1/2}.
\end{equation}

To form a prior distribution on $\bzeta$, we find the mean and covariance of the $\bzeta_{\text{opt}}$ values calculated as in Equation~\ref{eq:zeta-opt}.~Because of the choice of $\balpha$ and $\bF$ made above, the mean is zero and the covariance is diagonal with elements given by the eigenvalues of $\bS$.\footnote{We are guaranteed that these eigenvalues are non-negative because $\bS$ is a positive semi-definite matrix, as are all matrices of the form $\bXi\bXi^T$. Generally, we choose the number of basis vectors $n_b$ such that none of these eigenvalues are not too small.} We assume a prior distribution that is Gaussian with this mean and covariance.

\end{document}